\documentclass[aps,preprint,amsmath,amssymb,showpacs]{revtex4}
\usepackage{graphicx}
\usepackage{epstopdf}
\usepackage{color}
\begin{document}

\title{Search for the anomalous electromagnetic moments of tau lepton through electron-photon scattering at CLIC }

\author{Y.Ozguven}
\email[]{phyozguvenyucel@gmail.com} \affiliation{Department of
Physics, Cumhuriyet University, 58140, Sivas, Turkey}

\author{A. A. Billur}
\email[]{abillur@cumhuriyet.edu.tr} \affiliation{Department of
Physics, Cumhuriyet University, 58140, Sivas, Turkey}

\author{S.C. \.{I}nan}
\email[]{sceminan@cumhuriyet.edu.tr}
\affiliation{Department of Physics, Cumhuriyet University,
58140, Sivas, Turkey}

\author{M.K.Bahar}
\email[]{mussiv58@gmail.com} \affiliation{Department of Energy Systems Engineering, Karamanoglu Mehmetbey University, 70100, Karaman, Turkey}

\author{M. K\"{o}ksal}
\email[]{mkoksal@cumhuriyet.edu.tr} \affiliation{Department of Optical Engineering, Cumhuriyet University, 58140, Sivas, Turkey}

\begin{abstract}
We have examined the anomalous electromagnetic moments of the tau lepton in the processes $e^{-}\gamma \to \nu_e\tau\bar{\nu}_\tau$ ($\gamma$ is the Compton backscattering photon) and $e^{-}e^{+} \to e^{-}\gamma^* e^{+} \to \nu_{e}\tau \bar{\nu}_\tau e^+$ ($\gamma^*$
is the Weizsacker-Williams photon) with unpolarized and polarized electron beams at the CLIC. We have obtained 95$\%$ confidence level bounds on the anomalous magnetic and electric dipole moments for various values of the integrated luminosity and center-of-mass energy. Improved constraints of the anomalous magnetic and electric dipole moments have been obtained compared to the LEP sensitivity.
\end{abstract}


\maketitle

\section{Introduction}

The magnetic dipole moment of a particle is given by $ \vec{\mu}=g\left( e \hslash/2 m c\right)\vec{s} $ \cite{ger,uhlen}. Here, $g$ represents the strength of the magnetic dipole moment in units of Bohr magneton, and defined as g-factor or gyromagnetic factor. The value of $g$ for a point-like particle is obtained 2 as a result of the Dirac equation.
However, in quantum electrodynamics, the interactions of particle are much more complex and so there is a deviation from $ g=2 $ \cite{schwinger}. This deviation is known as anomalous magnetic moment. For any spin-$ 1/2 $ particle with mass, the anomalous magnetic moment is represented as $ a=(g-2)/2 $. The anomalous magnetic moments of electron and muon can be constrained with high accuracy at low energy spin precession experiments. The latest experimental data for the anomalous magnetic moment of the electron has been found as $ a_{e}=0.001159652180273(28) $\cite{hanna,data,berin}. Anomalous magnetic moment prediction can also be performed for the muon which has a mass about $ 207 $ times the electron mass. However, a disagreement has been observed for different Standard Model (SM) predictions and experimentally performed measurements for $ a_{\mu} $ \cite{hagi,bennet}. The latest experimental data  has been determined as $ a_{\mu}=0.00116592091(54)(53) $ through the $ E821 $ experiment \cite{hoec}. The experimental measurement of the anomalous magnetic moment of any particle contains its estimated value and some new physics effects that are just unpredictable in the SM. To find out the new physics contributions, it is an advantage that the mass of the tau lepton is enormous compared to the mass of the muon. However, the lifetime of the tau lepton is very short so measuring electric and magnetic dipole moments of the tau lepton is quite difficult with spin precession experiments. As a result, using colliders to study the anomalous magnetic moment of the tau lepton is highly preferable.

For the SM predictions, following numerical values can be found by summing all contributions discussed above \cite{11,12,hamzeh,samu}:

\begin{eqnarray}
a_{\tau}^{QED}=117324 \times 10^{-8}
\end{eqnarray}
\begin{eqnarray}
a_{\tau}^{EW}=47 \times 10^{-8}
\end{eqnarray}
\begin{eqnarray}
a_{\tau}^{HAD}=350.1 \times 10^{-8}
\end{eqnarray}
\begin{eqnarray}
a_{\tau}^{SM}=117721\times 10^{-8}=0.001177.
\end{eqnarray}

However, experimental restrictions on $ a_{\tau} $ have been obtained through $ e^{+}e^{-}\rightarrow e^{+}e^{-}\tau^{+}\tau^{-} $ by measuring the total cross-section at the $95\%$ C. L. in LEP in the following ranges \cite{L3,opal,del}:

\begin{center}
L3: $-0.052<a_{\tau}<0.058$, \\
OPAL: $-0.068<a_{\tau}<0.065$,\\
DELPHI: $-0.052<a_{\tau}<0.013$
\end{center}

The SM does not provide enough information to adequately understand the origin of CP violation \cite{chris}. Another interesting contribution in the interaction of photon with the tau lepton is CP violation which is generated by electric dipole moment. This phenomenon has been identified within the SM by the complex couplings in the CKM matrix of the quark sector \cite{koba}. In fact, there is no CP violation in the leptonic couplings (an exception is the neutrino mixing with different masses which is another source of CP violating \cite{barr}). Additional sources beyond the SM for the CP violation in the lepton sector are leptoquark \cite{18, 19}, SUSY \cite{20}, left-right symmetric \cite{21, 22} and Higgs models \cite{23, 24,25,26}. CP violation in the quark sector induces electric dipole moment of the leptons in the three loop level. Due to this contribution of the SM, it is very difficult to determine the electric dipole moment of the tau lepton. However, the electric dipole moment of this particle may cause detectable size due to interactions arising from the new physics beyond the SM.

The SM value for $d_{\tau}$ is obtained as $|d_{\tau}|\leq 10^{-34} e\,cm$ \cite{hoo}.
However, the most restrictive experimental bounds on the electric dipole moment $ d_{\tau} $  of the tau lepton have been obtained through $ e^{+}e^{-}\rightarrow\gamma\rightarrow\tau^{+}\tau^{-} $ by BELLE in the following ranges \cite{belle}:

\begin{center}
 $-2.2<Re(d_{\tau})<4.5 \times (10^{-17}\, e\,cm)$,\\
$-2.5<Im(d_{\tau})<0.8 \times (10^{-17}\, e\,cm)$.
\end{center}

 The main motivation of the present study is to investigate $ \tau\tau\gamma $ vertex contributions with anomalous electromagnetic form factors to the SM. In the SM, these form factors arise from radiative corrections. In this manner, to characterize the interaction of the tau lepton with the photon, the electromagnetic vertex factor can be parametrized by

\begin{eqnarray}
\Gamma^{\nu}=F_{1}(q^{2})\gamma^{\nu}+\frac{i}{2 m_{\tau}}F_{2}(q^{2}) \sigma^{\nu\mu}q_{\mu}+\frac{1}{2 m_{\tau}}F_{3}(q^{2}) \sigma^{\nu\mu}q_{\mu}\gamma^{5}
\end{eqnarray}
with $\sigma^{\nu\mu}=\frac{i}{2}(\gamma^{\nu}\gamma^{\mu}-\gamma^{\mu}\gamma^{\nu})$, where $ q $ and $ m_{\tau} $ are the photon momentum and the mass of the tau lepton, respectively. $ F_{1,2,3}\left( q^{2}\right)  $ are the electric charge, the anomalous magnetic dipole and the electric dipole form factors of the tau lepton. The electromagnetic form factor parametrizes $ \gamma^{\nu} $ electric charge coupling in the SM and electromagnetic coupling of the tau lepton for $ \tau\tau\gamma $ vertex is in compact form \cite{J,P,M}. There are a lot of phenomenological studies about this subject \cite{phe1,phe2,phe3,phe4,phe5,phe6,phe7}.

On the other hand, CLIC, aims to accelerate and collide electrons and positrons at $ 3 $ TeV nominal energy. It is a linear collider with high energy and high luminosity that is planned to be constructed at future date \cite{braun,clc}. In addition, CLIC can be constructed with $\gamma\gamma$ and $e \gamma$ collider modes with real photons. This real photon beam is obtained by the Compton backscattering of laser photons off linear electron beam.
Moreover, most of these photon beams can be in the high-energy region.

Linear colliders make it possible to use $\gamma^{*}\gamma^{*}$ and $e \gamma^{*}$ interactions possible to examine the new physics beyond the SM. The emitted photons from the incoming electrons scatter at very small angels from the
beam pipe. Therefore, these photons have very low virtuality and we say that these photons are
”almost-real”. The Weizs\"acker-Williams approximation is a facility in phenomenological studies because it permits to obtain cross sections for the process $e^{-}\gamma^{*} \rightarrow X $ approximately through the study of the main $e^{-}e^{+}\rightarrow e^{-}\gamma^{*} e^{+} \rightarrow Xe^{+}$ process.
Here, X represents particles obtained in the final state. Also, these interactions have very clean experimental conditions.

With these motivations, we have obtained the sensitivity bounds on new physics parameters through $e^- \gamma \to \nu_{e}\tau\bar{\nu}_\tau$ ($\gamma$ is the Compton backscattering photon) and  $e^{-}e^{+} \to e^{-}\gamma^* e^{+} \to \nu_{e}\tau\bar{\nu}_\tau e^{+}$ ($\gamma^*$ is Weizs\"acker-Williams photon) in next subsections.
In the next section, we briefly outline details of our numerical calculation and results. The final section is devoted to our conclusions.

\section{Numerical Analysis}

 In the SM,  electromagnetic form factors are reduced to $ F_{1}=1, F_{2}=F_{3}=0 $. However, due to the $ \tau\tau\gamma $ vertex, in other words, contributions from loop effects or arising from the new physics, $ F_{2} $ and $ F_{3} $ could not be taken as zero \cite{J,P,M}. While considering the limit of $ q^{2}\rightarrow0 $, the form factors become

\begin{eqnarray}
F_{1}(0)=1,\: F_{2}(0)=a_{\tau},\: F_{3}(0)=\frac{2m_{\tau}d_{\tau}}{e}
\end{eqnarray}
which relates to the static properties of fermions \cite{pich}.

\begin{figure}
\includegraphics{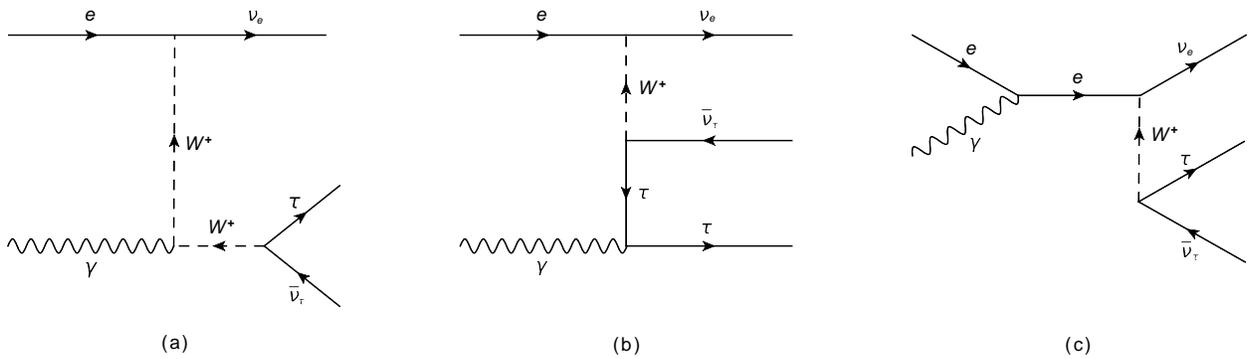}
\caption{Feynman diagrams of the $e^{-}\gamma \to \nu_e\tau\bar{\nu}_\tau$ subprocess.
\label{feyn}}
\end{figure}

In this study, the validity of the approach can be easily understood. As shown in the Feynman diagrams in Fig.\ref{feyn}, the anomalous electromagnetic
moments contribution of the tau lepton only comes from the diagram (b). As seen from this diagram, the photon in the $\gamma(\gamma^*)\tau\tau$ vertex is
either a Compton backscattering photon for the process $e^{-}\gamma \to \nu_{e}\tau \bar{\nu}_\tau$ or a Weizsacker-Williams photon for the $e^{-}e^{+} \to
e^{-}\gamma^* e^{+} \to \nu_{e}\tau\bar{\nu}_\tau e^{+}$ process. There is no other intermediate photon and Compton backscattering photon is on the mass-shell
($q^2=0$). So, the limit we use ($q^2 \to 0$) is appropriate for this process.

In Weizs\"acker-Williams approximation beam particles (electrons) scatter at very small angels.
So, electrons may not be observed in the central detector. If scattered electrons of the beams are detected, maximum and minimum values of incoming photon energies can be detected. In the other case, final energy or momentum cuts of produced final state particles can be used to specify minimum photon energy.
In this approximation, the photon virtuality is given by,

\begin{eqnarray}
\label{q2a}
Q^2=Q_{min}^2+\frac{q_t^2}{1-x}.
\end{eqnarray}

Here $x=E_\gamma/E$ is the ratio of the energy of the photon and the energy of the incoming electron, $Q^2=-q^2$ and $q_t$ is the transverse momentum of the photon.
$Q_{min}^2$ is given by

\begin{eqnarray}
\label{qmins}
Q_{min}^2=\frac{m_e^2 x^2}{1-x}.
\end{eqnarray}

\noindent  $Q_{min}^2$ is very small due to the electron mass (See Eq.\ref{qmins}).
In addition, since the electrons are scattered at very small angles, their transverse momentum are very small. For this reason, transverse momentum of the
emitted photons must be very small due to momentum conservation.
When all these arguments are taken into account, it can be understood that the virtuality of the photons in Weizs\"acker-Williams approximation should be
small. In other words, the photon must be almost-real.
The moment of the tau lepton was also investigated by the DELPHI collaboration using multiperipheral collisions through the process $e^+e^- \to e^+ e^-
\tau^+ \tau^-$  \cite{del}. In this study, the virtuality of $90\%$ of the photons was obtained as $1$ GeV$^2$  using the appropriate experimental
techniques. In this motivation, we have taken the maximum photon virtuality $2$ GeV${^2}$ as in other phenomenological studies.

In our calculations in this article, we have used the following kinematic cuts,

\begin{eqnarray}
p_{T}^{\nu_{e},\bar{\nu\tau}}&&>10 GeV, \nonumber \\
p_{T}^{\tau}&&>20 GeV,  \nonumber \\
|\eta_{\tau}|&&<2.5.
\end{eqnarray}

In sensitivity analysis, we take into account $\chi^2$ method,

\begin{eqnarray}
\chi^{2}=\left(\frac{\sigma_{SM}-\sigma(F_{2},F_{3})}{\sigma_{SM}\delta}\right)^{2},
\end{eqnarray}

\noindent where $\sigma(F_{2}, F_{3})$ is the total cross section which includes SM and new physics, $\delta=\sqrt{(\delta_{st})^{2}+(\delta_{sys})^{2}}$; $\delta_{st}=\frac{1}{\sqrt{N_{SM}}}$ is the statistical error and $\delta_{sys}$ is the
systematic error.

Systematic errors can arise for the following reasons.
First, we take into account experimental uncertainties.
However, we do not know exactly what the value of systematic errors of the two processes are since they are not examined in any of the CLIC reports
\cite{6,7,8}. The DELPHI collaboration examined the anomalous magnetic and electric dipole moments of the tau lepton through the process $e^{+}e^{-} \rightarrow
e^{+}e^{-}\tau^{+}\tau^{-}$ in the years $1997-2000$
at collision energy $\sqrt{s}$ between $183$ and $208$ GeV \cite{del}. Relative systematic errors on cross-section of the process $e^{+}e^{-} \rightarrow
e^{+}e^{-}\tau^{+}\tau^{-}$ are given in Table \ref{tabex}. Also, at center-of-mass energies $161$ GeV$\leqslant \sqrt{s} \leqslant 209$ GeV, the process $e^{+}e^{-}
\rightarrow e^{+}e^{-}\tau^{+}\tau^{-}$ was studied with the L3 detector at LEP \cite{L3}. The total systematic uncertainty in this work was estimated between $7\%$
and $9\%$. Even though the process $p p \rightarrow p p\tau^{+}\tau^{-}$ at the LHC has not been examined experimentally, the process $p p \rightarrow p
p\mu^{+}\mu^{-} $ at $\sqrt{s}=7$ TeV has been reported using data corresponding to an integrated luminosity of $40$ $pb^{-1}$. An overall relative systematic
uncertainty on the signal have obtained $4.8\%$ by summing quadratically all uncorrelated contributions \cite{10}. In addition, the anomalous magnetic and
electric dipole moments of the tau lepton via the process $p p \rightarrow p p\tau^{+}\tau^{-}$ with $2\%$ of the total systematic errors at the LHC was
investigated phenomenologically in Ref \cite{3}. As a result, we think that the systematic error in CLIC will be much smaller than these experimental studies
because it will be a new generation accelerator with innovative technologies.

Secondly,
there may be uncertainties arising from the identification of tau lepton. The tau lepton has several different decay channels. These channels
are classified according to the number of charged particles in the last state: one prong
or three prong. Since the particles in the tau decays are always greater than one, these
are called tau jets. The determination of hadronic decay channels are more problematic
than the leptonic modes due to QCD backgrounds. For hadronic decays tau jets can be
separated from other jets due to the its topology. Work in this regard is done by ATLAS
and CMS groups \cite{atlastau1, atlastau2, cmstau}. Tau tagging efficiencies also studied for ILD
\cite{ild}. Due to these difficulties, tau identification efficiencies are always calculated for specific process, luminosity, and kinematic parameters. These studies are currently being carried out by various
groups for selected productions. For a realistic efficiency, we need a detailed study for our
specific process and kinematic parameters. For all these reasons, in this work, kinematic
cuts contain some general values chosen by detectors for lepton identification. Hence, in this paper, tau lepton
identification efficiency is considered within systematic errors.

Thirdly, there may be theoretical uncertainties. One of these uncertainties may arise from photon spectra.
Another theoretical uncertainty comes from loop calculations in the SM, at tree level, $F_{1}=1$, $F_{2}=0$ and $F_{3}=0$. Besides, in the
loop effects arising from the SM and the new physics, $F_{2}$ and $F_{3}$ may not be equal to zero. For example, the anomalous coupling $F_{2}$ is given
by

\begin{eqnarray}
F_{2}(0)=a_{\tau}^{SM}+a_{\tau}^{NP}
\end{eqnarray}

\noindent where $a_{\tau}^{SM}$ is the contribution of the SM and $a_{\tau}^{NP}$ is the contribution of the new physics \cite{eidel, c1,c2,L3,c4}.
As mentioned above, $a_{\tau}^{SM}$ is the SM prediction comes from three parts (which occur the SM loop effects). Higher order corrections to the anomalous magnetic moment for tau lepton are searched for several authors in the literature. The total error on tau lepton anomalous magnetic moment which comes from QED, electroweak and hadronic loop contributions is approximately $ \delta=5.10^{-8}$ \cite{eidel, Roberts:2010zz, Passera:2004bj,Passera:2006gc, Fael:2014vyp}. This value is negligible compared to standard model value due to this reason not included uncertainty calculations. As a result,
we take into account the SM loop effects by using the electromagnetic vertex factor of the tau lepton.

We assume that the tau lepton decays into hadrons hence we take $BR=0.65$ in all calculations.
While getting the bound in this article, when fitting for $a_{\tau}$
we set $d_{\tau}$ to the SM value (zero), and \textit{vice versa}.

\subsection{ Analysis with Compton Backscattering Photons}

In this subsection, we show the numerical results for the $e^{-}\gamma \to \nu_{e}\tau \bar{\nu}_\tau$. We have used the CalcHEP package \cite{calchep,48} for all numerical analysis.  This program allows automatic
calculations of the distributions and cross sections in the SM as well as their extensions at the tree level. We have considered $\sqrt{s}=1.4$ TeV and $3$ TeV CLIC center-of-mass energies in our calculations.

The photon distribution function for the Compton backscattering photons is given by,

\begin{eqnarray}
f(x)=\frac{1}{g(\zeta)}\left[1-x+\frac{1}{1-x}-
\frac{4x}{\zeta(1-x)}+\frac{4x^{2}}{\zeta^{2}(1-x)^{2}}\right] ,
\end{eqnarray}
\noindent where
\begin{eqnarray}
g(\zeta)=\left(1-\frac{4}{\zeta}-\frac{8}{\zeta^2}\right)\log{(\zeta+1)}+
\frac{1}{2}+\frac{8}{\zeta}-\frac{1}{2(\zeta+1)^2} ,
\end{eqnarray}
\noindent with
\begin{eqnarray}
x=\frac{E_{\gamma}}{E_{e}} , \;\;\;\; \zeta=\frac{4E_{0}E_{e}}{m_{e}^2}
,\;\;\;\; x_{max}=\frac{\zeta}{1+\zeta} .
\end{eqnarray}

\noindent Here, $E_{0}$ and $E_{e}$ are  energy of the
incoming laser photon and initial energy of the electron beam before
Compton backscattering. $E_{\gamma}$ is the energy of the backscattered
photon. The maximum value of $x$ reaches $0.83$ when $\zeta=4.8$.

Using the above function, the cross section can be obtained as

\begin{eqnarray}
\label{cs}
 d\sigma=\int_{x_{min}}^{x_{max}}f(x)d\hat{\sigma}(\hat{s})
\end{eqnarray}
\noindent with $x_{min}=m_\tau^2/s$. Here $\hat{s}$ is related to $s$, the square of the center of mass energy of $e^-e^+$ collision, by $\hat{s}=xs$.

\begin{figure}
\includegraphics{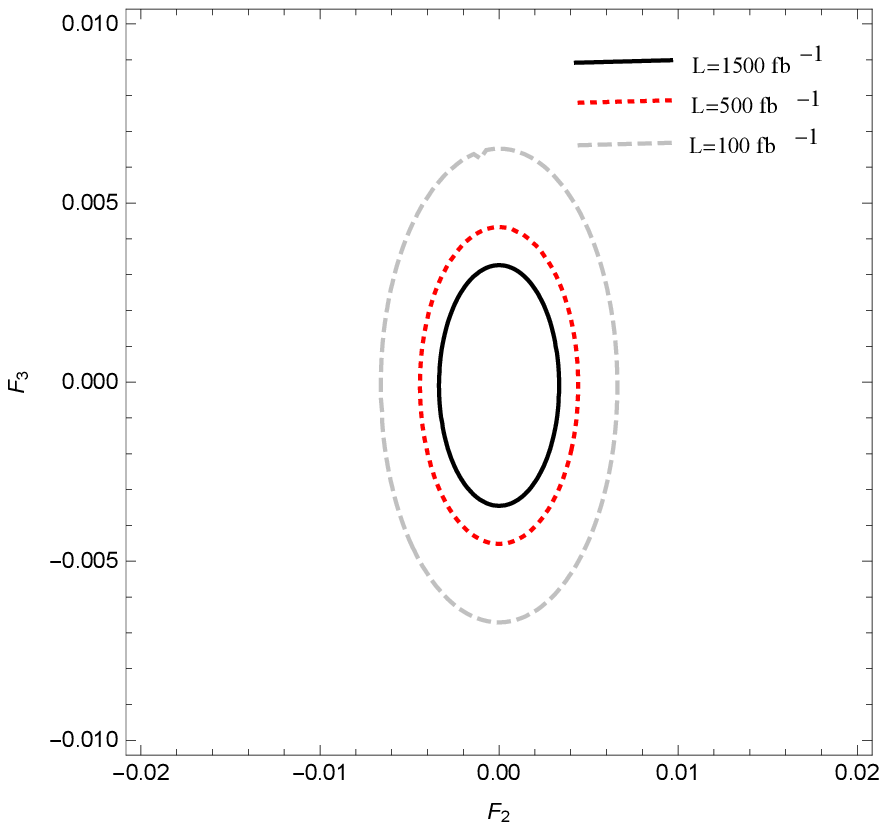}
\caption{Contour limits at the $95 \%$ C.L. in the $F_{2}-F_{3}$ plane for Compton backscattered photon $P_{e^-}=0\%$ and $\sqrt{s}=1.4$ TeV.
\label{fig1}}
\end{figure}

\begin{figure}
\includegraphics{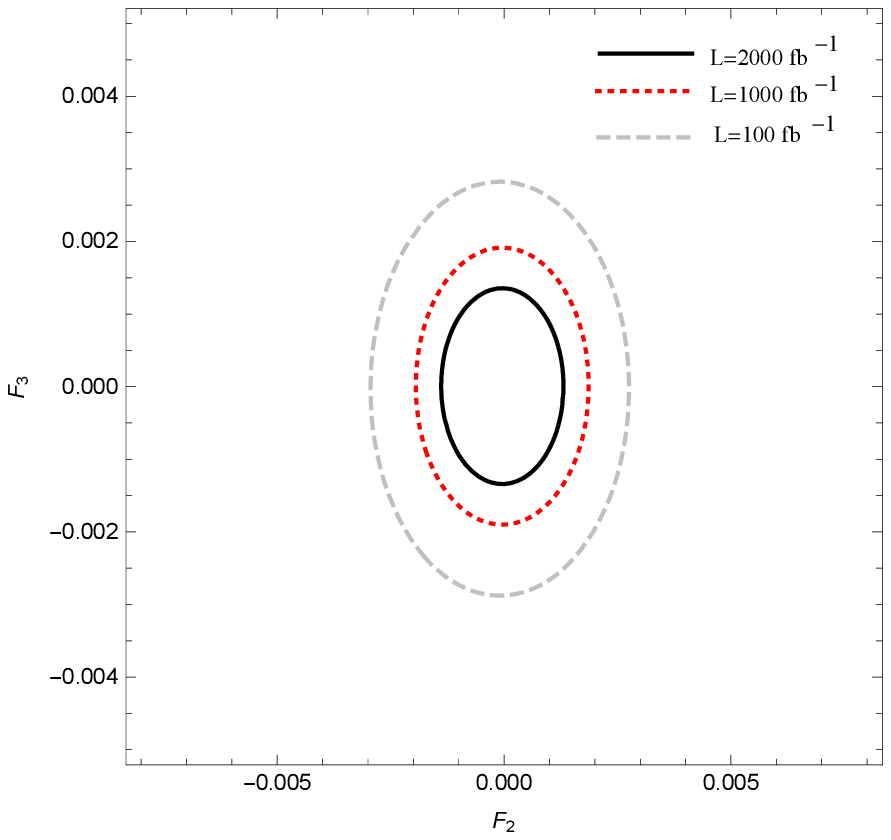}
\caption{Contour limits at the $95 \%$ C.L. in the $F_{2}-F_{3}$ plane for Compton backscattered photon $P_{e^-}=0\%$ and $\sqrt{s}=3$ TeV.
\label{fig2}}
\end{figure}

In Table \ref{tab1}, we present the $95\%$ C.L. sensitivity bounds on the anomalous couplings
for Compton backscattered photon and unpolarized electron beam ($P_{e^{-}}=0\%$), $\sqrt{s}=1.4$ TeV and $\sqrt{s}=3$ TeV center-of-mass energies and integrated CLIC luminosities. The bounds are found with no systematic error ($0\%$) and with systematic errors of $3\%$, $5\%$ $7\%$. Similarly, the limits on $\vert d_{\tau} \vert$ are shown in Table \ref{tab2}. It can be understood that the bounds on the anomalous couplings are sensitive to the values of the center-of-mass energy and luminosity. Also, we can see from these tables that our bounds on the $a_{\tau}$  are better than the current experimental limits even for $L=10$ fb$^{-1}$ and $\sqrt{s}=1.4$ TeV. In Figs. \ref{fig1} and \ref{fig2}, we show the contour bounds in the plane $F_{2}-F_{3}$ for $\sqrt{s}=1.4$ TeV with $L=100, 500, 1500$ fb$^{-1}$ and $\sqrt{s}=3$ TeV with $L=100, 1000, 2000$ fb$^{-1}$, respectively. The region outside the resulting ellipsoid are the regions of exclusion. From these figures, the best bounds on anomalous couplings are obtained for the $\sqrt{s}=3$ TeV and $L=2000$ fb$^{-1}$.

\begin{figure}
\includegraphics{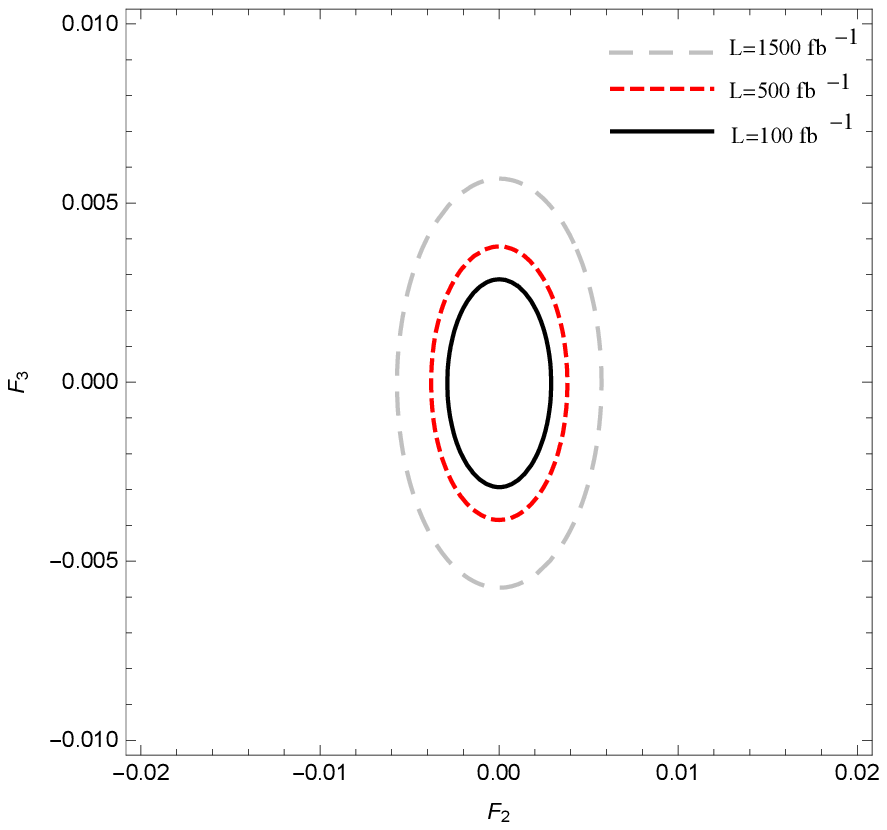}
\caption{Contour limits at the $95 \%$ C.L. in the $F_{2}-F_{3}$ plane for Compton backscattered photon $P_{e^-}=-80\%$ and $\sqrt{s}=1.4$ TeV.
\label{fig3}}
\end{figure}

\begin{figure}
\includegraphics{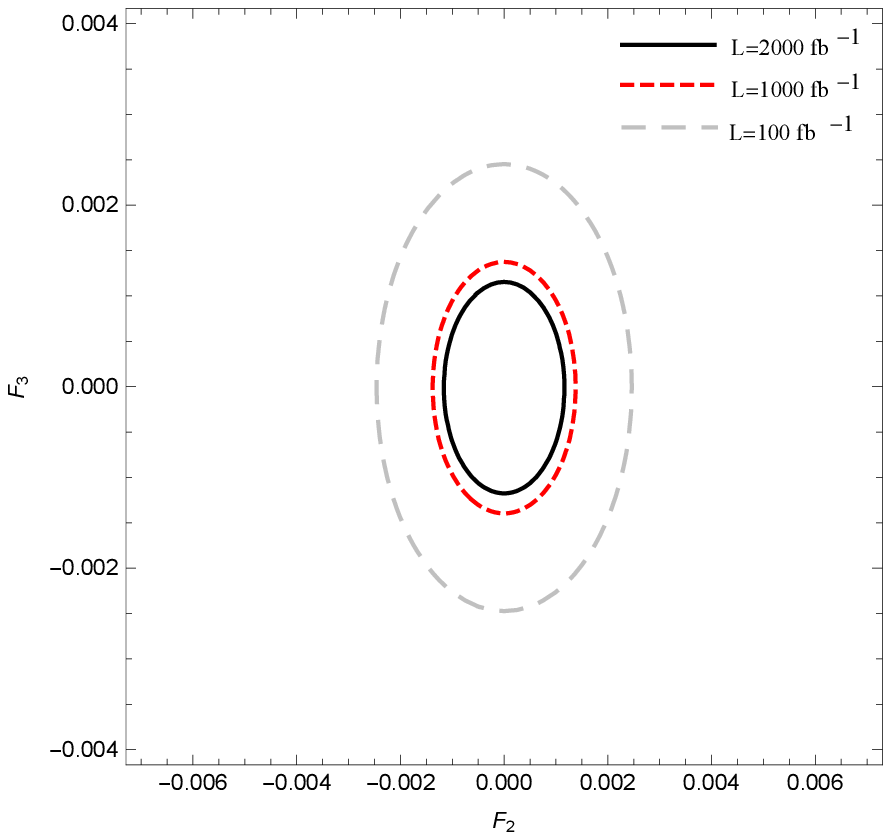}
\caption{Contour limits at the $95 \%$ C.L. in the $F_{2}-F_{3}$ plane for Compton backscattered photon $P_{e^-}=-80\%$ and $\sqrt{s}=3$ TeV.
\label{fig4}}
\end{figure}

For the numerical analysis, we have used polarized electron beams. For a process with electron and positron beam polarizations, the cross section can be defined as \cite{pol},

\begin{eqnarray}
\sigma=\frac{1}{4}(1-P_{e^+})(1+P_{e^-})\sigma_{-1+1}+\frac{1}{4}(1+P_{e^+})(1-P_{e^-})\sigma_{+1-1}.
\end{eqnarray}

\noindent where $\sigma_{ab}$ represents the obtained cross section with fixed helicities $a$ for positron and $b$ for the electron.
$P_{e^-}$ and  $P_{e^+}$ are the polarization degree of the electron and positron, respectively.
The process which is examined in this paper, has three Feynman diagrams each of them have weak charged boson vertex. Due to the
weak bosons couple to left handed fermions, negative helicity polarization can increase cross section and as a consequence of the increment in the cross
section, the stronger bounds on the anomalous electromagnetic moments can be achieved.
Hence, we have applied $ -80\% $ $P_{e^-}$ electron polarization. We give $95\%$ C.L. sensitivity bounds on the anomalous $a_{\tau}$ and $\vert d_{\tau} \vert$ couplings in Tables \ref{tab3} and  \ref{tab4} respectively. Here, we have
considered $P_{e^-}=-80\%$  for $\sqrt{s}=1.4$ and $\sqrt{s}=3$ TeV with different luminosity values. As seen from the tables obtained sensitivity bounds on the anomalous couplings are better than unpolarized beams.  In Figs. \ref{fig3} and \ref{fig4} for polarized electron beam, we show the contour bounds in the plane $F_2-F_3$ for $\sqrt{s}=1.4$ TeV with $L=100, 500, 1500$ fb$^{-1}$ and $\sqrt{s}=3$ TeV with $L=100, 1000, 2000$ fb$^{-1}$, respectively. Comparison of Figs. \ref{fig1} (\ref{fig2}) and \ref{fig3} (\ref{fig4}) shows that the
excluded area of the model $F_2-F_3$ parameters which we have obtained from the polarized beams ($P_{e^-}=-80\%$) expands to wider regions than the cases of the unpolarized electron beams.

\subsection{ Analysis with Weizsacker-Williams Photons}

We have analyzed the anomalous dipole moments of the tau lepton via the main process $e^{-}e^{+} \to e^{-}\gamma^* e^{+} \to \nu_{e}\tau\bar{\nu}_\tau e^{+}$ in this subsection. In Weizsacker-Williams approximation, the photon spectrum used in the CalcHep
program is 

\begin{eqnarray}
\label{epa}
\frac{dN}{dE_\gamma}=f(x)=\frac{\alpha}{\pi
E_e}\left[\left(\frac{1-x+x^2/2}{x}\right)\log\frac{Q_{max}^2}{Q_{min}^2}-\frac{m_e^2x}{Q_{min}^2}\left(1-\frac{Q_{min}^2}{Q_{max}^2}\right)\right]
\end{eqnarray}

\noindent where $m_{e}$ is the mass of the electron, $Q^2=-q^2$,  $x=E_\gamma/E_e$ is the ratio of the energy of the photon and energy of the incoming electron, $\alpha=1/137.035$ is the fine structure constant. Using Eq.(\ref{epa}), the cross section can be obtained by using Eq.\ref{cs}

 In Tables \ref{tab5} and \ref{tab6} we present the $95\%$ C.L. sensitivity bounds on the anomalous  $a_{\tau}$ and  $\vert d_{\tau} \vert$ parameters for the unpolarized electron beams and different systematic error values, respectively. We can understand from the tables that the sensitivity bounds of the anomalous couplings enhance with the increasing center-of-mass energy and luminosity. The obtained bounds for the $a_{\tau}$ are also better than the current experimental limits. On the other hand, bounds with a Compton backscattering photon (Tables \ref{tab1} and \ref{tab2} ) are more sensitive than the bounds for the Weizs\"acker-Williams approximation (Table \ref{tab5} and \ref{tab6}). Main reason of this situation is that the Compton backscattering photon spectrum gives higher effective than the Weizs\"acker-Williams photon spectrum in high energy regions \cite{com1, com2,ep1,ep2,ep3,ep4,ep5,wwa}. However, the application of the Weizs\"acker-Williams approximation gives a lot of benefits in experimental and phenomenological studies \cite{ph1,ph2,ph3,ph4,ph5,ph6,ph7,qm1,qm2,qm3,qm4} as mentioned in Section I. We present the $95\%$ C.L. sensitivity bounds on the anomalous $a_{\tau}$ and $\vert d_{\tau} \vert$  parameters for the $P_{e^-}=-80\%$ polarized electron beams in Table \ref{tab7} and \ref{tab8} for $\sqrt{s}=1.4$ TeV with $L=10, 100, 500, 1500$ fb$^{-1}$ and $\sqrt{s}=3$ TeV with $L=10, 500, 1000, 2000$ fb$^{-1}$, respectively. Best bounds on the anomalous couplings have been obtained in this situation as we expected due to above discussions.

\begin{figure}
\includegraphics{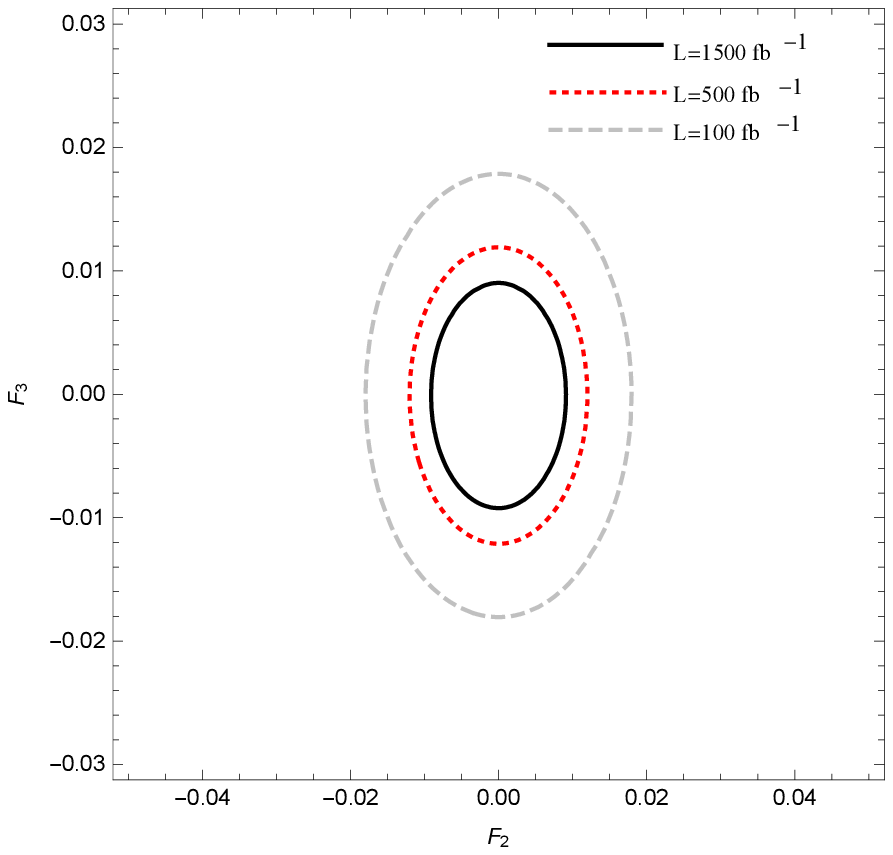}
\caption{Contour limits at the $95 \%$ C.L. in the $F_{2}-F_{3}$ plane for Weizsacker-Williams photon $P_{e^-}=0\%$ and $\sqrt{s}=1.4$ TeV.
\label{fig5}}
\end{figure}

\begin{figure}
\includegraphics{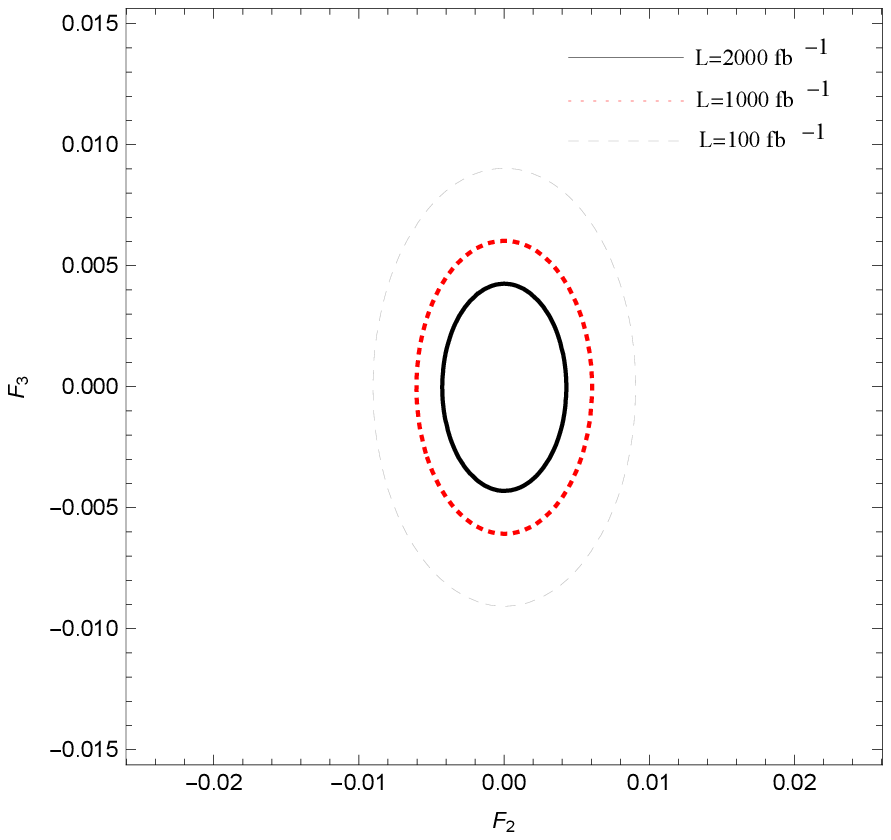}
\caption{Contour limits at the $95 \%$ C.L. in the $F_{2}-F_{3}$ plane for Weizsacker-Williams photon $P_{e^-}=0\%$ and $\sqrt{s}=3$ TeV.
\label{fig6}}
\end{figure}

\begin{figure}
\includegraphics{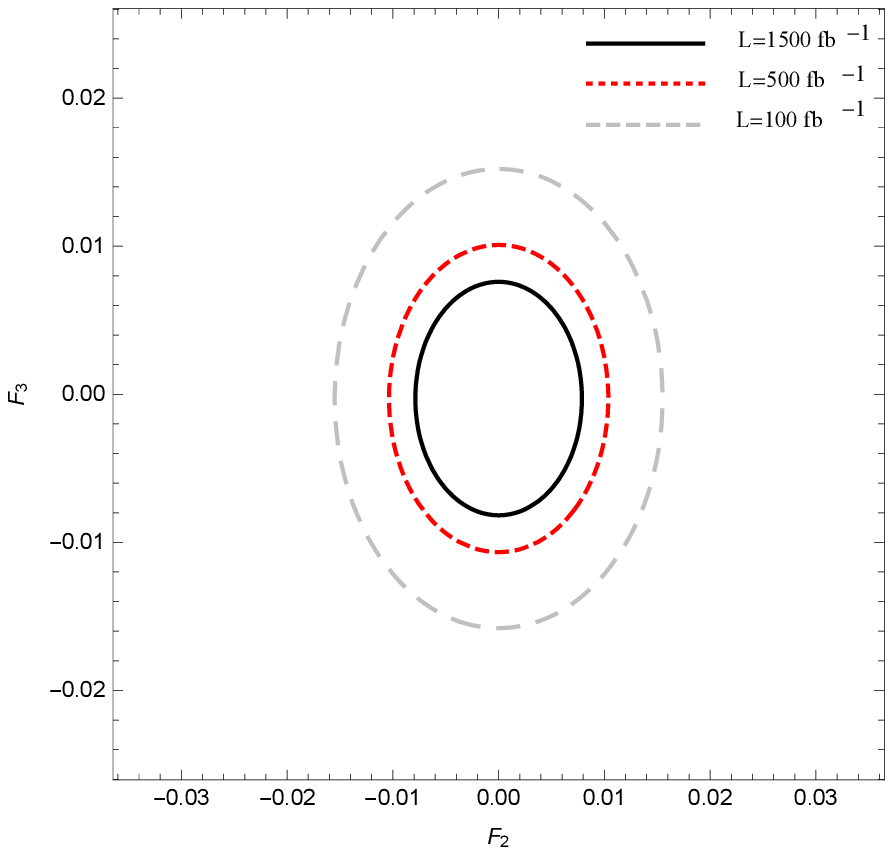}
\caption{Limits contours at the $95 \%$ C.L. in the $F_{2}-F_{3}$ plane for Weizsacker-Williams photon $P_{e^-}=-80\%$ and $\sqrt{s}=1.4$ TeV.
\label{fig7}}
\end{figure}

\begin{figure}
\includegraphics{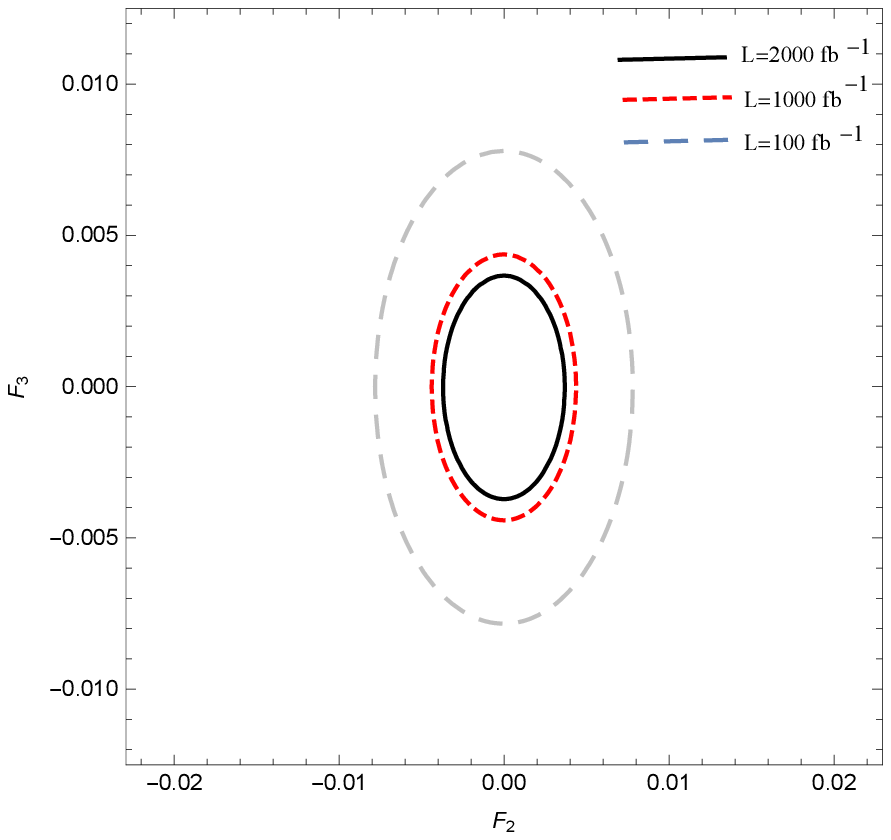}
\caption{Limits contours at the $95 \%$ C.L. in the $F_{2}-F_{3}$ plane for Weizsacker-Williams photon $P_{e^-}=-80\%$ and $\sqrt{s}=3$ TeV.
\label{fig8}}
\end{figure}

In Figs. \ref{fig5} and \ref{fig6} for unpolarized electron beam we present the contour bounds in the plane $F_2-F_3$ for $\sqrt{s}=1.4$ TeV with $L=100, 500, 1500$ fb$^{-1}$ and $\sqrt{s}=3$ TeV with $L=100, 1000, 2000$ fb$^{-1}$, respectively . Fig.\ref{fig7} (Fig.\ref{fig8}) is the same as Fig. \ref{fig5} (Fig.\ref{fig6}) but for polarized electron beams. Limits for the polarized case are strong compared to the unpolarized case. However, these bounds are weaker compared to the Compton backscattered case.

\section{Conclusion}

We analyzed the tau lepton anomalous dipole moments through the processes $e^- \gamma  \to \nu_{e}\tau \bar{\nu}_\tau$ and  $e^{-}e^{+} \to e^{-}\gamma^* e^{+} \to \nu_{e}\tau\bar{\nu}_\tau e^{+}$. These processes have a very clean environment. The deviation of the anomalous couplings
from the expected values of the SM would evidence the existence of new physics. In this study, we compared the electromagnetic dipole moments of the tau lepton using the Weizs\"acker approximation and Compton back-scattering photons. We have found the $e^- \gamma  \to \nu_{e}\tau\bar{\nu_{\tau}}$  process gives better bounds than the other. However, processes that have $\gamma\gamma$ and $e^- \gamma $ initial states require a special collider setup. On the other hand, $e^{-}\gamma^*$ and $\gamma^* \gamma^*$ occur spontaneously during $ e^{+}e^{-}  $  collisions.

Additionally, we used polarized and unpolarized electron beam in our study. We understood the polarization enhances the sensitivity bounds as mentioned in Section II. Our predictions for the expected limits on $a_{\tau}$ are better than the current experimental limits. Based on the finding of this paper, we can conclude that CLIC provides new opportunities for examination of tau physics beyond the SM using $e\gamma$ and $e\gamma^*$ modes.

\section{Acknowledgements}
This work has been supported by the Scientific and Technological Research Council of Turkey (TUBITAK) in the framework of Project No. 115F136.

\begin{table}
\caption{Systematic errors given by the DELPHI collaboration \cite{del}.
\label{tabex}}
\begin{ruledtabular}
\begin{tabular}{ccccc}
 & $1997$& $1998$& $1999$& $2000$ \\
\hline
Trigger efficiency& $7.0$& $2.7$& $3.6$& $4.5$ \\
Selection efficiency& $5.1$& $3.2$& $3.0$& $3.0$  \\
Background& $1.7$& $0.9$& $0.9$& $0.9$  \\
Luminosity& $0.6$& $0.6$& $0.6$& $0.6$ \\
Total& $8.9$& $4.3$& $4.7$& $5.4$  \\
\end{tabular}
\end{ruledtabular}
\end{table}

\begin{table}
\caption{ 95\% C.L. sensitivity bounds of the $a_{\tau}$ couplings
for Compton backscattered photon and unpolarized electron beam, various center-of-mass energies and integrated CLIC
luminosities.  The bounds are showed with no systematic error ($0\%$) and with systematic errors of $3\%$, $5\%$ $7\%$.\label{tab1}}
\begin{ruledtabular}
\begin{tabular}{cccccc}
 $\sqrt{s}$ TeV & Luminosity($fb^{-1}$)& $0\%$ & $3\%$ & $5\%$ & $7\%$  \\
\hline
    &   10 & (-0.0105, 0.0105) &(-0.0227, 0.0227) & (-0.0291, 0.0291) &(-0.0343, 0.0343)  \\
    &  100 & (-0.0059, 0.0059) &(-0.0224, 0.0225) & -0.0290, 0.0290) &(-0.0342, 0.0343)  \\
1.4 &  500 & (-0.0039, 0.0039) &(-0.0224, 0.0224 & (-0.0289, 0.0290) &(-0.0342, 0.0342)  \\
    & 1500 & (-0.0030, 0.0030) & (-0.0224, 0.0224)& (-0.0289, 0.0290) &(-0.0342, 0.0342)  \\
\hline
    &   10 & (-0.0046, 0.0046) &(-0.0092, 0.0092) & (-0.0118, 0.0118) &(-0.0139, 0.0139)  \\
    &  500 & (-0.0017, 0.0017) &(-0.0091, 0.0091) & (-0.0117, 0.0117) &(-0.0139, 0.0139)  \\
3   & 1000 & (-0.0014, 0.0014) &(-0.0091, 0.0091) & (-0.0117, 0.0117) &(-0.0139, 0.0139)  \\
    & 2000 & (-0.0012, 0.0012) &(-0.0091, 0.0091) & (-0.0117, 0.0117) &(-0.0139, 0.0139)  \\
\end{tabular}
\end{ruledtabular}
\end{table}

\begin{table}
\caption{Same as the Table \ref{tab1} but for the $\vert d_{\tau} \vert$.\label{tab2}}
\begin{ruledtabular}
\begin{tabular}{cccccc}
 $\sqrt{s}$ TeV & Luminosity($fb^{-1}$)& $0\%$ & $3\%$ & $5\%$ & $7\%$  \\
\hline
    &   10 & $ 0.59\times 10^{-15}$ & $ 1.26\times 10^{-15}$ & $ 1.61\times 10^{-15}$ & $  1.91\times 10^{-15}$ \\
    &  100 & $ 0.33\times 10^{-15}$ &$ 1.25\times 10^{-15}$ & $ 1.61\times 10^{-15}$ & $  1.90\times 10^{-15}$ \\
1.4 &  500 & $ 0.22\times 10^{-15}$ & $ 1.24\times 10^{-15}$ & $ 1.61\times 10^{-15}$ & $  1.90\times 10^{-15}$ \\
    & 1500 & $ 0.17\times 10^{-15}$ & $ 1.24\times 10^{-15}$ & $ 1.61\times 10^{-15}$ & $  1.90\times 10^{-15}$ \\
\hline
    &   10 & $ 0.26\times 10^{-15}$ & $ 0.51\times 10^{-15}$ & $ 0.65\times 10^{-15}$ & $  0.77\times 10^{-15}$  \\
    &  500 & $ 0.09\times 10^{-15}$ & $ 0.50\times 10^{-15}$ & $ 0.65\times 10^{-15}$ & $  0.77\times 10^{-15}$  \\
3   & 1000 & $ 0.08\times 10^{-15}$ & $ 0.50\times 10^{-15}$ & $ 0.65\times 10^{-15}$ & $  0.77\times 10^{-15}$  \\
    & 2000 & $ 0.07\times 10^{-15}$ & $ 0.50\times 10^{-15}$ & $ 0.65\times 10^{-15}$ & $  0.77\times 10^{-15}$  \\
\end{tabular}
\end{ruledtabular}
\end{table}

\begin{table}
\caption{ 95\% C.L. sensitivity bounds of the $a_{\tau}$ couplings
for Compton backscattered photon and $-80\%$ polarized electron beam, various center-of-mass energies and integrated CLIC
luminosities.  The bounds are showed with no systematic error ($0\%$) and with systematic errors of $3\%$, $5\%$ $7\%$. \label{tab3}}
\begin{ruledtabular}
\begin{tabular}{cccccc}
 $\sqrt{s}$ TeV & Luminosity($fb^{-1}$)& $0\%$ & $3\%$ & $5\%$ & $7\%$  \\
\hline
    &   10 & (-0.0091, 0.0091) & (-0.0226, 0.0225) & (-0.0290, 0.0290)& (-0.0343, 0.0342) \\
    &  100 & (-0.0051, 0.0051) & (-0.0225, 0.0224) & (-0.0290, 0.0289) &(-0.0343, 0.0342)\\
1.4 &  500 & (-0.0034, 0.0034) & (-0.0224, 0.0224) & (-0.0290, 0.0289) &(-0.0343, 0.0342) \\
    & 1500 & (-0.0026, 0.0026) & (-0.0224, 0.0224) & (-0.0290, 0.0289) &(-0.0343, 0.0342) \\
\hline
    &   10 & (-0.0039, 0.0039) & (-0.0091, 0.0091) & (-0.0117, 0.0117) &(-0.0138, 0.0138) \\
    &  500 & (-0.0015, 0.0015) & (-0.0090, 0.0091) & (-0.0117, 0.0117) &(-0.0138, 0.0138) \\
3   & 1000 & (-0.0012, 0.0012) & (-0.0090, 0.0091) & (-0.0117, 0.0117) &(-0.0138, 0.0138) \\
    & 2000 & (-0.0010, 0.0010) & (-0.0090, 0.0091) & (-0.0117, 0.0117) &(-0.0138, 0.0138) \\
\end{tabular}
\end{ruledtabular}
\end{table}

\begin{table}
\caption{ Same as the Table \ref{tab3} but for the  $\vert d_{\tau} \vert$.\label{tab4}}
\begin{ruledtabular}
\begin{tabular}{cccccc}
 $\sqrt{s}$ TeV & Luminosity($fb^{-1}$) & $0\%$ & $3\%$ & $5\%$ & $7\%$  \\
\hline
    &   10 & $ 0.51\times 10^{-15}$ & $ 1.25\times 10^{-15}$ & $ 1.61\times 10^{-15}$ & $  1.90\times 10^{-15}$\\
    &  100 & $ 0.28\times 10^{-15}$ & $ 1.25\times 10^{-15}$ & $ 1.61\times 10^{-15}$& $  1.90\times 10^{-15}$\\
1.4 &  500 & $ 0.19\times 10^{-15}$ & $ 1.24\times 10^{-15}$ & $ 1.61\times 10^{-15}$&$  1.90\times 10^{-15}$ \\
    & 1500 & $ 0.14\times 10^{-15}$ & $ 1.24\times 10^{-15}$ & $ 1.61\times 10^{-15}$ &$  1.90\times 10^{-15}$ \\
\hline
    &   10 & $ 0.22\times 10^{-15}$ & $ 0.51\times 10^{-15}$ & $ 0.65\times 10^{-15}$ & $  0.77\times 10^{-15}$\\
    &  500 & $ 0.08\times 10^{-15}$ & $ 0.50\times 10^{-15}$ & $ 0.65\times 10^{-15}$ &$  0.77\times 10^{-15}$\\
3   & 1000 & $ 0.07\times 10^{-15}$ & $ 0.50\times 10^{-15}$ & $ 0.65\times 10^{-15}$ &$  0.77\times 10^{-15}$\\
    & 2000 & $ 0.06\times 10^{-15}$ & $ 0.50\times 10^{-15}$ & $ 0.65\times 10^{-15}$ &$  0.77\times 10^{-15}$ \\
\end{tabular}
\end{ruledtabular}
\end{table}

\begin{table}
\caption{95\% C.L. sensitivity bounds of the $a_{\tau}$ couplings
for Weizsacker-Williams photon and unpolarized electron beam, various center-of-mass energies and integrated CLIC
luminosities. The bounds are showed with no systematic error ($0\%$) and with systematic errors of $3\%$, $5\%$ $7\%$. \label{tab5}}
\begin{ruledtabular}
\begin{tabular}{cccccc}
 $\sqrt{s}$ TeV & Luminosity($fb^{-1}$)& $0\%$ & $3\%$ & $5\%$ & $7\%$  \\
\hline
    &   10 & (-0.0287, 0.0285) & (-0.0404, 0.0402) & (-0.0499, 0.0496) &(-0.0582, 0.0579) \\
    &  100 & (-0.0162, 0.0160) & (-0.0379, 0.0376) & (-0.0486, 0.0483) &(-0.0574, 0.0571) \\
1.4 &  500 & (-0.0109, 0.0106) & (-0.0376, 0.0374) & (-0.0485, 0.0482) &(-0.0573, 0.0571) \\
    & 1500 & (-0.0083, 0.0081) & (-0.0376, 0.0373) & (-0.0484, 0.0482) &(-0.0573, 0.0571) \\
\hline
    &   10 & (-0.0144, 0.0144) & (-0.0221, 0.0220) & (-0.0276, 0.0275) &(-0.0323, 0.0323) \\
    &  500 & (-0.0054, 0.0054) & (-0.0210, 0.0209) & (-0.0271, 0.0270) &(-0.0320, 0.0320) \\
3   & 1000 & (-0.0046, 0.0045) & (-0.0210, 0.0209) & (-0.0271, 0.0270) &(-0.0320, 0.0320) \\
    & 2000 & (-0.0039, 0.0038) & (-0.0210, 0.0209) & (-0.0271, 0.0270) &(-0.0320, 0.0320) \\
\end{tabular}
\end{ruledtabular}
\end{table}

\

\begin{table}
\caption{ Same as the Table \ref{tab5} but for the $\vert d_{\tau} \vert$. \label{tab6}}
\begin{ruledtabular}
\begin{tabular}{cccccc}
 $\sqrt{s}$ TeV & Luminosity($fb^{-1}$) & $0\%$ & $3\%$ & $5\%$ & $7\%$  \\
\hline
    &   10 & $ 1.57\times 10^{-15}$ & $ 2.24\times 10^{-15}$ & $ 2.76\times 10^{-15}$&$  3.22\times 10^{-15}$ \\
    &  100 & $ 0.87\times 10^{-15}$ & $ 2.10\times 10^{-15}$ & $ 2.69\times 10^{-15}$ &$  3.18\times 10^{-15}$ \\
1.4 &  500 & $ 0.56\times 10^{-15}$ & $ 2.08\times 10^{-15}$ & $ 2.68\times 10^{-15}$ &$  3.17\times 10^{-15}$ \\
    & 1500 & $ 0.41\times 10^{-15}$ & $ 2.08\times 10^{-15}$ & $ 2.68\times 10^{-15}$ &$  3.17\times 10^{-15}$ \\
\hline
    &   10 & $ 0.79\times 10^{-15}$ & $ 1.22\times 10^{-15}$ & $ 1.53\times 10^{-15}$ & $  1.79\times 10^{-15}$\\
    &  500 & $ 0.29\times 10^{-15}$ & $ 1.17\times 10^{-15}$ & $ 1.50\times 10^{-15}$ & $  1.78\times 10^{-15}$\\
3   & 1000 & $ 0.26\times 10^{-15}$ & $ 1.16\times 10^{-15}$ & $ 1.50\times 10^{-15}$ &$  1.78\times 10^{-15}$ \\
    & 2000 & $ 0.21\times 10^{-15}$ & $ 1.16\times 10^{-15}$ & $ 1.50\times 10^{-15}$ &$  1.78\times 10^{-15}$ \\
\end{tabular}
\end{ruledtabular}
\end{table}

\begin{table}
\caption{95\% C.L. sensitivity bounds of the $a_{\tau}$ couplings
for Weizsacker-Williams photon and $-80\%$ polarized electron beam, various center-of-mass energies and integrated CLIC
luminosities. The bounds are showed with no systematic error ($0\%$) and with systematic errors of $3\%$, $5\%$ $7\%$.\label{tab7}}
\begin{ruledtabular}
\begin{tabular}{cccccc}
 $\sqrt{s}$ TeV & Luminosity($fb^{-1}$) & $0\%$ & $3\%$ & $5\%$ & $7\%$  \\
\hline
    &   10 & (-0.0248, 0.0246) & (-0.0392, 0.0390) & (-0.0493, 0.0490) &(-0.0578, 0.0575) \\
    &  100 & (-0.0140, 0.0138) & (-0.0377, 0.0375) & (-0.0485, 0.0483) &(-0.0574, 0.0571)  \\
1.4 &  500 & (-0.0094, 0.0092) & (-0.0376, 0.0373) & (-0.0485, 0.0482) &(-0.0573, 0.0571) \\
    & 1500 & (-0.0072, 0.0069) & (-0.0376, 0.0373) & (-0.0485, 0.0482) & (-0.0573, 0.0571)\\
\hline
    &   10 & (-0.0124, 0.0124) & (-0.0216, 0.0215) & (-0.0274, 0.0273) &(-0.0322, 0.0322) \\
    &  500 & (-0.0046, 0.0046) & (-0.0210, 0.0209) & (-0.0271, 0.0270) &(-0.0320, 0.0320) \\
3   & 1000 & (-0.0039, 0.0039) & (-0.0210, 0.0209) & (-0.0271, 0.0270) &(-0.0320, 0.0320) \\
    & 2000 & (-0.0033, 0.0033) & (-0.0210, 0.0209) & (-0.0271, 0.0270) &(-0.0320, 0.0320) \\
\end{tabular}
\end{ruledtabular}
\end{table}

\begin{table}
\caption{  Same as the Table \ref{tab7} but for the $\vert d_{\tau} \vert$.\label{tab8}}
\begin{ruledtabular}
\begin{tabular}{cccccc}
 $\sqrt{s}$ TeV & Luminosity($fb^{-1}$) & $0\%$ & $3\%$ & $5\%$ & $7\%$  \\
\hline
    &   10 & $ 1.37\times 10^{-15}$ & $ 2.17\times 10^{-15}$ & $ 2.73\times 10^{-15}$ &$  3.20\times 10^{-15}$ \\
    &  100 & $ 0.77\times 10^{-15}$ & $ 2.09\times 10^{-15}$ & $ 2.69\times 10^{-15}$ &$  3.18\times 10^{-15}$ \\
1.4 &  500 & $ 0.52\times 10^{-15}$ & $ 2.08\times 10^{-15}$ & $ 2.68\times 10^{-15}$ &$  3.17\times 10^{-15}$ \\
    & 1500 & $ 0.38\times 10^{-15}$ & $ 2.08\times 10^{-15}$ & $ 2.68\times 10^{-15}$ &$  3.17\times 10^{-15}$ \\
\hline
    &   10 & $ 0.68\times 10^{-15}$ & $ 1.20\times 10^{-15}$ & $ 1.52\times 10^{-15}$ &$  1.79\times 10^{-15}$ \\
    &  500 & $ 0.26\times 10^{-15}$ & $ 1.16\times 10^{-15}$ & $ 1.50\times 10^{-15}$ &$  1.78\times 10^{-15}$\\
3   & 1000 & $ 0.22\times 10^{-15}$ & $ 1.16\times 10^{-15}$ & $ 1.50\times 10^{-15}$ &$  1.78\times 10^{-15}$ \\
    & 2000 & $ 0.18\times 10^{-15}$ & $ 1.16\times 10^{-15}$ & $ 1.50\times 10^{-15}$ &$  1.78\times 10^{-15}$ \\
\end{tabular}
\end{ruledtabular}
\end{table}

\end{document}